# Basic Insights into Tunable Graphene Hydrogenation

Ricarda A. Schäfer, Daniela Dasler, Udo Mundloch, Frank Hauke and Andreas Hirsch*

Department of Chemistry and Pharmacy and Joint Institute of Advanced Materials and Processes (ZMP), Friedrich-Alexander University of Erlangen-Nürnberg, Henkestrasse 42, 91054 Erlangen (Germany)

**ABSTRACT:** The hydrogenation and deuteration of graphite with potassium intercalation compounds (GICs) as start-ing materials was investigated. Characterization of the reactions products (hydrogenated and deuterated graphene) was carried out by thermogravimetric analysis coupled with mass spectrometry (TG-MS) and Raman spectroscopy in-cluding statistical Raman spectroscopy (SRS) and –microscopy (SRM). The results reveal that the choice of the hydro-gen/deuterium source, the nature of the graphite (used as starting material), the potassium concentration in the inter-calation compound as well as the choice of the solvent have a great impact on the reaction outcome. Furthermore, it was possible to proof that both mono and few-layer hydrogenated/deuterated graphene can be produced.

**Introduction:** The synthesis and complete characterization of fully hydrogenated graphene (graph**a**ne) remains to be a challenge in carbon allotrope chemistry.[1] Graphane has been theoretically studied by Sofo *et al.*[2] Recently, a variety of reports towards the synthesis of partially hydrogenated graphene have been published. These include the hydrogenation of graphene using hydrogen plasma[3] as well as wet chemical approaches.[4,5] In the latter case, the synthetic protocols are based on a Birch[6] type reduction sequence utilizing alcohols or water as proton sources. Liquid ammonia and alkali metals, like lithium or sodium, served as reducing agents. At the same time the reduction of graphite facilitates the exfoliation of graphene sheets driven by the electrostatic repulsion of the charged individual layers. We have demonstrated that the application of water as proton source leads to very high degrees of hydrogenation and the resulting stable reaction product shows a remarkable fluorescence behaviour.[5] A critical drawback of this approach however is a) the use of the somewhat difficult to handle liquid ammonia and b) the restriction to control the degree of hydrogenation. We now introduce a new route towards hydrogenated graphene using potassium intercalated graphite as starting materials and THF as inert solvent (Scheme 1). This reductive graphite activation procedure has already proven to be a versatile concept for the bulk synthesis of covalently functionalized graphene as the intermediate graphenides can be trapped by a variety of electrophiles like diazonium[7] or iodonium compounds[8] and alkyl or aryl iodides.[9,10] As a consequence, we have expected that reduced graphite intercalation compounds (GICs) are also suitable starting materials for subsequent protonations leading to the formation of hydrogenated graphene. Since in GICs the degree of reduction can easily be controlled by the stoichiometry of the alkali metal it should be possible to control the extent of hydrogenation, which is another advantage of this approach.

In order to study the general feasibility of the GIC protonation, we have systematically screened three different types of graphite - synthetic spherical (SGN18), natural flake (NG), and expanded powder (PEX10) - and varied a) the amount of potassium and b) the nature of the proton/deuteron source in order to obtain deeper insights into the underlying reactivity principles.

The reaction products have been characterized in detail by scanning Raman spectroscopy (SRS) and –microscopy (SRM)[9] as well as thermogravimetric analysis coupled with mass spectrometry (TG-MS). Our results clearly show that in this way the bulk synthesis of hydrogenated graphene - with tuneable hydrogen content - can be accomplished.

In previous studies it has been noticed that the combination of $KC_8$ with proton sources leads to the formation of varying products ranging from nanoscrolls[11] to partly hydrated or hydrogenated graphite depending on the reaction conditions.[12] We have recently shown that the ultrasound agitation of potassium intercalated graphite leads to an efficient exfoliation of the graphitic starting material and to the generation of negatively charged monolayer graphene.[7] The corresponding experimental setup has been adapted in the present study.

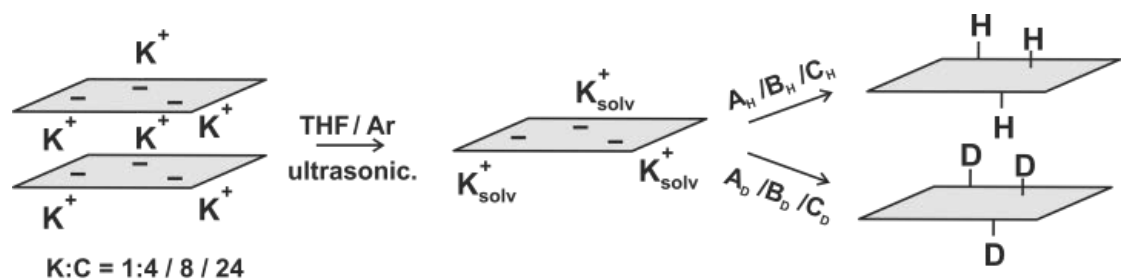

**Scheme 1.** Hydrogenation of graphene using three types of graphite (synthetic spherical graphite (SGN18), natural flake graphite (NG), and expanded powder graphite (PEX10)) as starting materials. In the initially generated potassium GICs the K:C ratio has been varied (K:C = 1:4, 1:8, 1:24). Three different sources for protons ($A_H=H_2O$, $B_H=MeOH$, $C_H=t$-BuOH) and deuterons ($A_D=D_2O$, $B_D=MeOD$, $C_D=t$-BuOD) have been used.

In particular, the potassium GIC (SGN 18 - spherical graphite, 18 μm flake size) has been exfoliated in THF *via* sonication and subsequently the intermediately generated graphenide flakes have been trapped with 10 eq. of a hydrogen/deuterium source. The dispersion of the intercalate in THF *via* sonication marks the fundamental difference to previously reported approaches,[11-14] which were carried out without ultrasound agitation and in the most cases in the absence of any solvent. The Coulomb force driven exfoliation of graphite has a fundamental impact on the reaction mechanism and on the nature of the final product, since the efficient individualization of the packed graphene layers provides a huge and highly reactive surface area for the subsequent covalent functionalization. In analogy to other reductive functionalization sequences, a single-electron transfer from the charged graphenide sheets to the trapping water/alcohol should lead to the formation of hydrogen radicals.[9] Subsequently, these radicals can either recombine to molecular hydrogen or attack the extended $sp^2$-carbon lattice yielding hydrogenated graphene sheets. We suggest that due to the close proximity between the generated radicals and the graphene sheets the latter process (hydrogenation) is predominant.

In order to prove that the successful hydrogenation of graphene is possible, thermogravimetric analysis coupled with mass spectrometry (TG-MS) has been applied. For this purpose the samples were heated from rt to 700 °C under a constant flow of helium and the thermally detached entities were analyzed by an EI mass spectrometer. As exemplarily depicted in Fig. 1 for **G$_{1:4}$C$_H$** (GIC potassium concentration of K:C=1:4 and *t*-BuOH as trapping reagent) two major steps of mass loss, namely region A: 220 to 350 °C and region B: 380 to 500 °C (indicated in red) – can be identified. The first step can be attributed to molecular fragments exhibiting $^m/_z$ 17/18/31 and 72 (Fig. S1), which are due to residues of water, hydroxides or THF that have been co-intercalated or encapsulated between the re-aggregated graphene sheets. These findings are in total agreement with previously published results.[8]

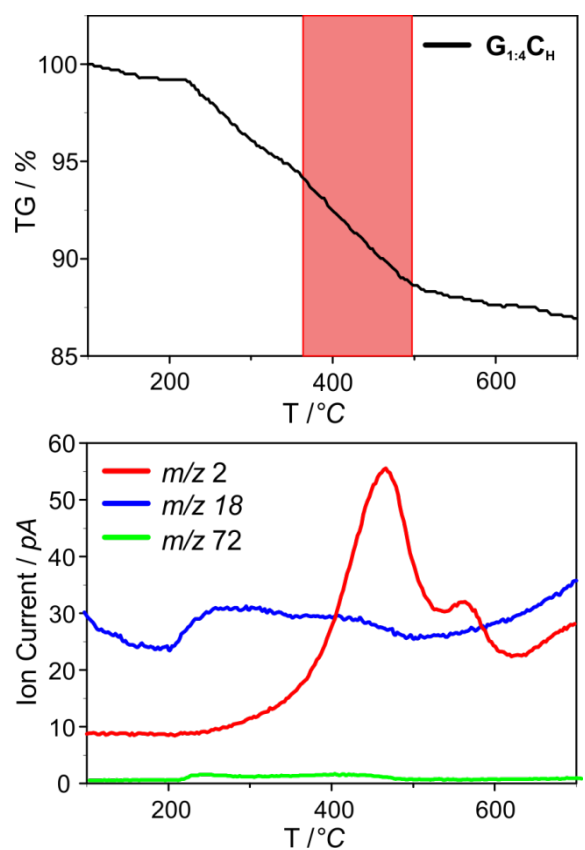

**Figure 1.** Top: TG profile of hydrogenated graphene **G$_{1:4}$C$_H$**. Bottom: MS trace for $^m/_z$ 2 (H$_2$), 18 (H$_2$O), and 72 (THF).

In the second region B, the observed mass loss of 4.8 % can directly be attributed to the dehydrogenation of the sample, as the mass spectrometric analysis exhibits a pronounced peak (60 pA intensity) for $^m/_z$ 2 in this temperature range. To ensure that the detected hydrogen is introduced by the alcohol trapping reagent we substituted the hydrogen source with a deuterium source. TG-MS analysis of the sample $G_{1:4}C_D$ has been carried out under a constant nitrogen flow, exhibiting a similar mass loss of 5.5 % (Fig. 2a). Simultaneously, the recorded molecular fragments with $^m/_z$ 3 and 4 (Fig. 2b and Fig. S2) can be attributed to DH and $D_2$ with a maximum ion current at the same temperature as measured for $G_{1:4}C_H$.

Thus, it can be concluded that the hydrogen or the deuterium is directly introduced by the addition of the respective trapping reagent. It is most likely that the appearance of HD can be traced back to hydrogen exchange reactions with the residual $H_2O$ traces in the THF solvent.

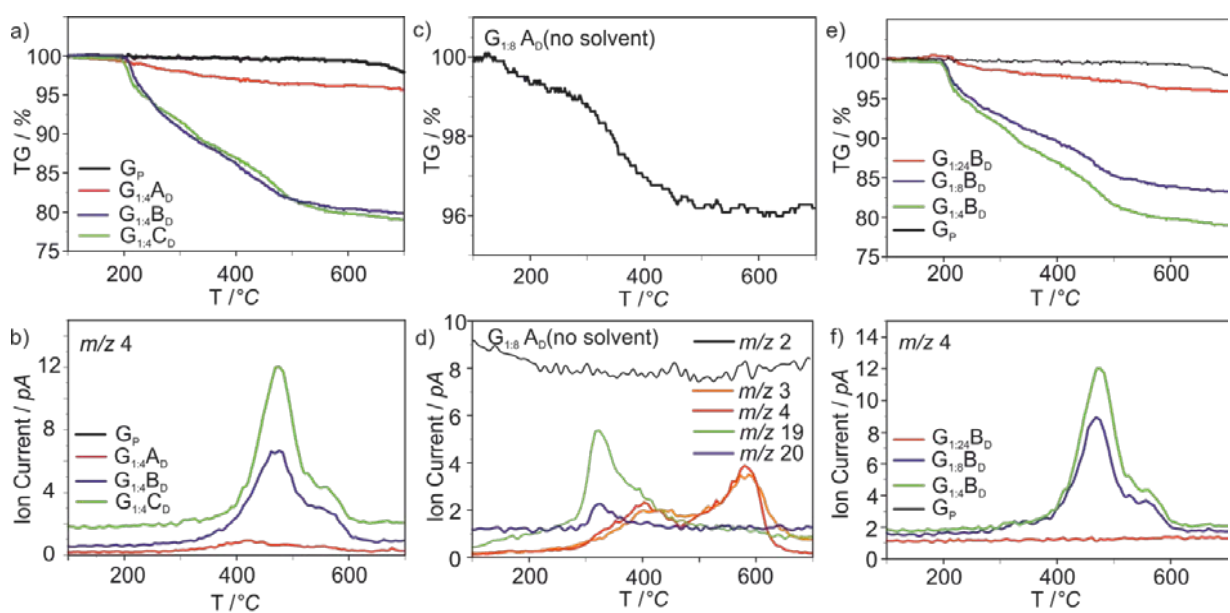

**Figure 2.** (a) TG profile of deuterated graphene $G_P$, $G_{1:4}A_D$, $G_{1:4}B_D$, and $G_{1:4}C_D$; (b) corresponding MS trace for $^m/_z$ 4 ($D_2$); (c) TG profile of $G_{1:4}A_D$(no solvent); (d) MS trace for $^m/_z$ 2 ($H_2$), 3 (HD), 4($D_2$), 19 (HDO), and 20 ($D_2O$); (e) TG profile of $G_{1:4}B_D$, $G_{1:8}B_D$, $G_{1:24}B_D$, and $G_P$; (f) MS trace for $^m/_z$ 4($D_2$) of $G_{1:4}B_D$, $G_{1:8}B_D$, and $G_{1:24}B_D$.

For a deeper understanding of the influence of the deuterium source we investigated three different systems, namely, *t*-BuOD, MeOD, and heavy water. The amount of deuteration can slightly be increased by the use of deutero-methanol (6.1 % mass loss in temperature region B) with respect to *t*-BuOD (5.5 % mass loss), but drastically decreases when $D_2O$ (2.2 % mass loss) was applied as trapping reagent (Fig. 2a,b). Interestingly, this reaction outcome is in contrast to the results obtained under classical Birch reduction conditions, where water yields hydrogenated graphene with the highest degree of functionalization.[5] This fact can be explained with the nature of the solvent. In the latter case, ammonia is used at -75 °C and interacts with the trapping water (formation of ammonium hydroxide in an equilibrium)

In order to check this rationale, we used liquid ammonia instead of THF for the exfoliation/dispersion of the potassium intercalation compound **G$_{1:4}$** and D$_2$O as trapping component (Scheme 2).

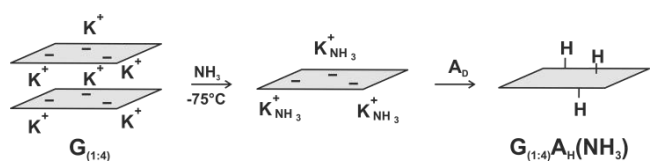

**Scheme 2:** Reaction of spherical graphite intercalated with potassium (K:C = 1:4). The intercalation compound was dispersed in liquid ammonia at -75 °C and the reaction mixture was quenched with D$_2$O.

The TG profile of the final product **G$_{1:4}$A$_D$(NH$_3$)** exhibits a drastically reduced mass loss in the temperature region from 100 to 350 °C in comparison to **G$_{1:4}$A$_D$**, indicative for the absence of an encapsulated species (Fig. S3). In addition, the maximum ion current for the mass trace $^m/_z$ 2 is detected at 450 °C. Due to deuterium/hydrogen exchange reactions no mass traces with $^m/_z$ 3 or 4 were detected. Furthermore, we could demonstrate the influence of the solvent and reproduce the experiment of Schlögl et al.[13] by adding D$_2$O to **G$_{1:8}$** in the total absence of any GIC exfoliating solvent. Here, a mass loss of < 1 % is detected in the temperature region B between 380 °C to 600 °C (Fig. 2c,d) and a small extent of deuteration ($^m/_z$ 3 and 4 max at 550 °C) can be verified by MS analysis. The predominant sample mass loss (2.4 %) takes place in the temperature region A (200 to 380 °C) where the mass traces $^m/_z$ 19 and 20 can be attributed to HDO and D$_2$O. The increased detachment temperature for the $^m/_z$ 3 and 4 fragments (550 °C vs. 450 °C) in the case of **G$_{1:4}$A$_D$** might be explained by a predominant edge functionalization due to an inefficient exfoliation of the charged GICs. In addition, the formation of HDO probably is based on the presence of partially encapsulated KOD which cannot be removed by washing and drying during the work-up (reference experiment with KOH see Fig. S4). These encapsulation scenario is in accordance to the findings of Schlögl et al.[13] The results of our reference experiments clearly underline the importance of the proper solvent for the reductive hydrogenation/deuteration of graphene.

In the next step we studied the impact of the potassium concentration on the hydrogenation efficiency. In the case of the reductive phenylation of graphene we have already demonstrated that the degree of reduction determines the degree of covalent addend binding.[8,15] As outlined in Scheme 1, we varied the amount of potassium (K:C = 1:4, 1:8, 1:24) by using MeOD as deuteration source.

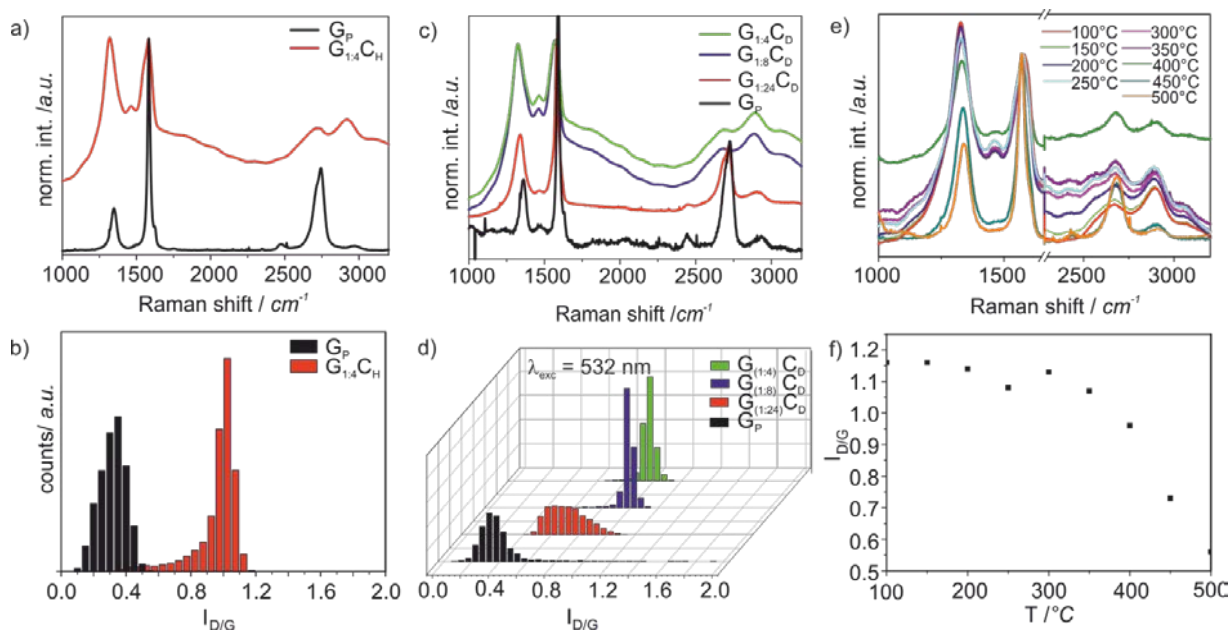

**Figure 3.** Statistical Raman spectroscopy of hydrogenated and deuterated graphene. (a) Raman mean spectrum of **G$_P$** and **G$_{1:4}$C$_H$**. (b) Statistical distribution (10.000 single point spectra) of the I$_{D/G}$ intensity ratio of the pristine starting material **G$_P$** (black) and the functionalized bulk material **G$_{1:4}$C$_H$** (red). (c) Raman mean spectrum of **G$_P$**, **G$_{1:4}$C$_D$**, **G$_{1:8}$C$_D$**, and **G$_{1:24}$C$_D$**. (d) Statistical distribution (10.000 single point spectra) of the I$_{D/G}$ intensity ratio of the pristine starting material **G$_P$** (black) and the functionalized bulk material **G$_{1:4}$C$_D$** (green), **G$_{1:8}$C$_D$** (blue), and **G$_{1:24}$C$_D$** (red). (e) Mean Raman spectra of **G$_{1:4}$B$_D$** at 100 °C – 500 °C. (f) I$_{D/G}$ ratio of **G$_{1:4}$B$_D$** at 100 °C – 500 °C.

The TG-MS data (Fig. 2e,f) for **G$_{1:8}$B$_D$** and **G$_{1:4}$B$_D$** demonstrate an increase of the detected mass loss from 4.9 % (0.125 equivalents K/C) to 6.1 % for 0.25 eq. K/C and a simultaneous ion current intensity increase for $^m/_z$ 4 (D$_2$) from 8.9 pA to 12 pA, whereas the decrease of potassium to 0.04 equivalents per carbon reduces the amount of deuteration almost to zero (Fig. 2e,f, Fig. S5-7). The results for the analogously hydrogenated samples are summarized in Fig. S8-10.

Taking into account that for graphane – fully hydrogenated graphene – a maximum mass loss of 7.7 % can be expected, our obtained 4.8 % in the case of the SGN 18 (synthetic spherical graphite) starting GIC, with 0.25 eq. potassium, represents a very promising starting point for future graphane based applications.

In addition, it can be expected - and it has been shown[8,15] - that the type of graphite, used as starting material in reductive functionalization sequences, has an impact on the final degree of functionalization. In order to investigate the basic influence of the morphology of the starting graphite we used a natural flake graphite (NG) and an expanded powder graphite (PEX10) in addition to the synthetic spherical graphite SGN18 in the course of the reductive deuteration with *t*-BuOD as trapping reagent.

In the case of the powder graphite (3-5 µm flake size) the respective TG profile of **G$_{1:4}$C$_D$(PEX)** (Fig. S11) exhibits only one region of mass loss between 300 and 500 °C ($^m/_z$ 2,

3 and 4 detected by MS). Interestingly, the smaller flake size efficiently inhibits the encapsulation of THF or potassium hydroxide.[8,13] Whereas a bigger flake size of 1,000 μm as present in the natural graphite leads to the opposite effect (increase of mass loss region A) and a less effective deuteration – 3.6 % mass loss in region B for **G$_{1:4}$C$_D$(PEX)** *vs.* 3.1 % mass loss for **G$_{1:4}$C$_D$(NG)** – Fig S11. As a consequence, it can be clearly seen that the hydrogenation of graphene can be influenced by the nature of the hydrogen/deuterium source, the nature of starting material, and the amount of intercalated potassium. The resulting material can reach a maximum of 4.8 % hydrogen mass loss (6.1 % deuterium mass loss) and is, in comparison to the polyhydrogenated graphene, conducting and non-fluorescent which can be attributed to a different hydrogenation pattern.

Further unambiguous proof for the successful hydrogenation/deuteration of graphene was provided by a detailed Raman spectroscopic characterization including statistical Raman spectroscopy (SRS) and statistical Raman microscopy (SRM).[16] In general, the Raman spectrum of graphene exhibits three main peaks - the defect induced D-band at 1,350 cm$^{-1}$, the G-band at 1,582 cm$^{-1}$, which is characteristic for the graphitic sp$^2$-carbon lattice, and the 2D-band at 2,700 cm$^{-1}$ providing information about the stacking or electronic individualization of the graphene layers.[17] The amount of basal plane sp$^3$-carbon atoms, bearing a hydrogen/deuterium addend, correlates with the $I_{D/G}$ intensity ratio in the Raman spectra and therefore corresponds to the degree of functionalization.[13,14,18] In Fig. 3a the Raman spectrum of the highly hydrogenated sample **G$_{1:4}$C$_H$** is presented as an example.

The statistical Raman analysis reveals a comparatively narrow distribution of basal plane sp$^3$-centers in the functionalized bulk material **G$_{1:4}$C$_H$**, which is indicative for a highly homogeneously functionalized bulk material. A mean $I_{D/G}$ ratio of 1.0 (Fig 3b) together with a very broad D- and G-band and a fluorescence background is indicative for a covalently functionalized graphene derivative with a very high degree of functionalization. We have recently obtained similar data for hydrogenated graphene prepared under classical Birch conditions.[5] All these characteristic features can be taken as a clear indication for the introduction of an extensive amount of sp$^3$-lattice carbon atoms carrying the covalently bound addends.[5,9,13,14,18] An important conclusion which can be drawn from the results of the in-depth Raman characterization of the hydrogenated and deuterated samples is, that the reductive protonation/deuteration of intermediately generated graphenides yields covalently functionalized bulk materials with identical $I_{D/G}$ ratios (**G$_{1:4}$C$_H$** = 1.0, **G$_{1:4}$C$_D$** = 1.0 –Fig. 3b and Fig. 3d, respectively). Furthermore, the nature of the starting graphite – (SGN18, NG, PEX) has no pronounced effect on the final $I_{D/G}$ ratio (Fig. S12). In consistency with the TG-MS results, it turns out that the main parameter which allows for a fine-tuning of the degree of functionalization is the potassium concentration used in the respective GIC starting materials.

This is clarified in the following section for the reductive deuteration of SGN18 ($G_{1:n}$) – the complete data set deuteration/hydrogenation is presented in Fig. S13-18.

In the deuterated samples the mean $I_{D/G}$ ratio (Fig. 3c,d) for $G_{1:4}A_D$ is only slightly smaller (0.9) than the mean $I_{D/G}$ ratios of $G_{1:4}B_D$ and $G_{1:4}C_D$ (1.0) – again no significant influence of the nature of the starting graphite is observed. Samples with the lowest potassium concentration in the starting GIC ($G_{1:24}A$-$C$) yield the lowest $I_{D/G}$ intensity ratio (0.7), whereas for $G_{1:8}A_D$-$C_D$ an intermediate $I_{D/G}$ ratio of 0.9 is detected. Moreover, the potassium concentration has also a direct influence on the homogeneity of the final material (Fig. 3d & Fig. S13-18). Here, in both cases of reductive deuteration/hydrogenation all samples with $G_{1:24}$ potassium concentration yield a remarkably broadened $I_{D/G}$-ratio distribution, indicative for an in-homogeneously functionalized bulk material.

In analogy to the TG-MS experiments, the thermal cleavage of the hydrogen and deuterium addends can be directly monitored by Raman spectroscopy. For this purpose $G_{1:4}B_D$ was heated under an $N_2$ atmosphere from 100 to 500 °C and every 50 °C step a Raman map of 1,000 spectra was recorded. As presented in Fig. 3e, the mean spectra reveal a decreasing D-band intensity with increasing temperature.

Significantly, the plot of the $I_{D/G}$ values as function of the temperature (Fig. 3f) represents exactly the same trend as the corresponding TG profile. Consistently, a pronounced decrease of the detected $I_{D/G}$ band intensity is only observed at temperatures higher than 350 °C. This observation is in full agreement with the detected mass loss and recorded mass traces of H/D in the temperature region B (detach from 350 to 500 °C) and an unequivocal proof for covalently bound hydrogen/deuterium atoms.

The sample characterization presented so far was primarily based on the analysis of the bulk material. Here, we were able to demonstrate that an efficient bulk hydrogenation is possible on the basis of potassium intercalated graphite starting materials. That this reductive reaction protocol indeed leads to functionalized monolayer and few-layer graphene can be shown by a detailed investigation of drop-casted samples of the reaction mixture of $G_{1:4}C_D(NG)$. In Fig. 4 two characteristic SRM images of deuterated graphene flakes (denoted with 1 and 2) are presented.

A remarkable observation is, that the spectral data of the single flakes also confirm the results of the bulk measurement with respect to the obtained degree of functionalization ($I_{D/G}$ of 1.0 and a broadened D-band). Moreover, it clearly poofs the successful production of a deuterated mono-layer (flake 1) and few-layer graphene (flake 2) with a respective $I_{2D/G}$ ratio between 3.0 and 1.0 and a 2D-band FWHM of 30 cm$^{-1}$ for flake 1 and 59 cm$^{-1}$ for flake 2.

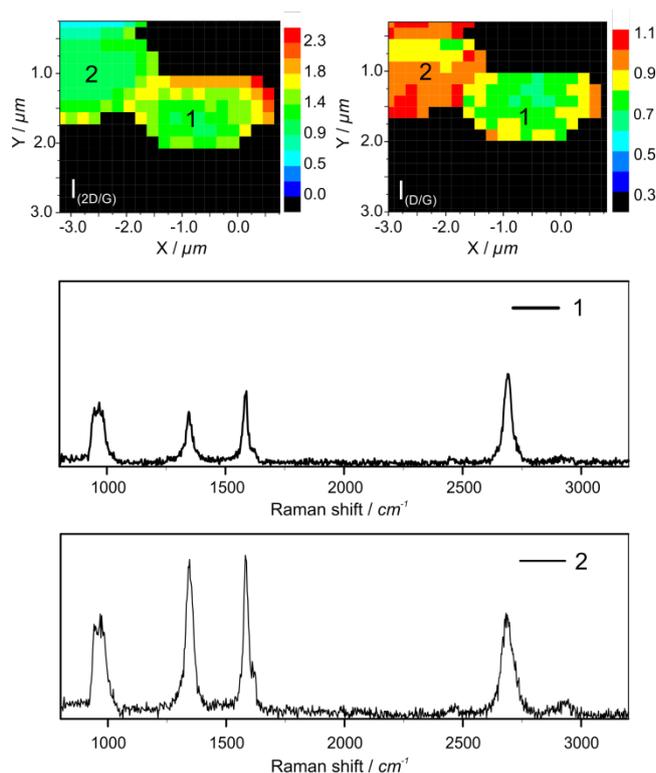

**Figure 4** Top: SRM image of **G$_{1:4}$C$_D$(NG)** flake 1 and 2 – left: plot of the I$_{2D/G}$ intensity ratio; right: plot of the I$_{D/G}$ intensity ratio. Bottom: Raman single-point spectra of flake 1 and flake 2.

In conclusion, we presented an efficient protocol for the reductive hydrogenation/deuteration of potassium intercalated graphite as starting material. In this way, a highly hydrogenated bulk material (4.8 % mass loss of hydrogen – 7.7 % theoretical value for graph**a**ne) is accessible. The successful synthesis has been proven by detailed TG-MS and Raman spectroscopic investigations. The variation of the graphite starting material, hydrogen/deuterium source, and potassium concentration showed that these parameters have a distinct influence on the hydrogen content in the final material.

By the right combination of these parameters, a tuneable degree of functionalization can be accomplished. In addition, the covalent binding of hydrogen is reversible and the attached hydrogen atoms can thermally be detached in a temperate region between 350 and 600 °C. As concluded from SRS and SRM, the reductive hydrogenation yields a functionalized bulk material mainly consisting of hydrogenated mono-layer and few-layer graphene.

## ACKNOWLEDGMENT

The authors thank the Deutsche Forschungsgemeinschaft (DFG- SFB 953, Project A1 "Synthetic Carbon Allotropes") for financial support. The research leading to these results has received partial funding from the European Union Seventh Framework Programme under grant agreement no.604391 Graphene Flagship.

# Basic Insights into Tunable Graphene Hydrogenation

# Electronic Supporting Information

*Ricarda A. Schäfer, Daniela Dasler, Udo Mundloch, Frank Hauke and Andreas Hirsch\**

## 1. Materials

Synthetic spherical graphite (SGN18, 99.99 % C, TGA residue 0.01 % wt - Future Carbon, Germany) with a mean grain size of 18 μm and a specific surface area of 6.2 m$^2$/g was used after annealing under vacuum (300 °C).

Natural flake graphite (NG, TGA residue 2.38 %, Kropfmühl AG, Germany) - with a mean grain size of ~1,000 μm and a specific surface area of 0.2 m$^2$/g was used as received.

Expanded powder graphite (PEX10, 98.5 % C, TGA residue 0.54 % wt - Future Carbon, Germany) with a mean grain size of 3-5 μm and a specific surface area of 6 m$^2$/g was used as received.

Chemicals and solvents were purchased from Sigma Aldrich Co. (Germany) and were used as-received if not stated otherwise.

THF was distilled in an argon inert gas atmosphere once over sodium and three times over sodium-potassium alloy in order to remove any remaining water. Residual traces of oxygen were removed by pump freeze treatment (four iterative steps). THF$_{(abs)}$ was used for all reactions.

The deuterated alcohols and heavy water were degased by pump freeze treatment (five iterative steps) to remove remaining traces of oxygen.

## 2. Equipment and Characterization

*Glove Box:* Sample synthesis and preparation was carried out in an argon filled LABmaster[pro] sp glove box (MBraun), equipped with a gas purifier and solvent vapor removal unit: oxygen content < 0.1 ppm, water content < 0.1 ppm.

*Raman Spectroscopy:* Raman spectroscopic characterization was carried out on a LabRAM Aramis confocal Raman microscope (Horiba) with a laser spot size of about 1 µm (Olympus LMPlanFI 50x LWD, NA 0.50) in backscattering geometry. As excitation source a green laser with $\lambda_{exc}$ = 532 nm was used and the incident laser power was kept as low as possible (1.35 mW) to avoid any structural sample damage. Spectra were recorded with a CCD array at -70 °C – grating: 600 grooves/mm. Spectra were obtained from a 50 µm x 50 µm area with 2 µm step size in SWIFT mode for low integration times. Exact sample movement was provided by an automated xy-scanning table. Calibration in frequency was carried out with a HOPG crystal as reference.

## Thermogravimetric Analysis (TGA) in combination with a mass spectrometer (MS): TG/MS Analysis

Mass spectrometer coupled thermogravimetric analysis (TG/MS) was carried out on a Netzsch STA 409 CD instrument equipped with a Skimmer QMS 422 mass spectrometer (EI ion source, quadrupole mass spectrometer) with the following programmed time-dependent temperature profile: 24 - 700 °C (dynamic heating of 10 K/min). Sample analysis (initial sample weight about 5 mg) was carried out under inert gas atmosphere with a helium gas flow of 80 mL/min. For the characterization of the deuterated graphene derivatives nitrogen was used as carrier gas.

## 3. Experimental

*Graphite intercalation compounds with varying potassium:carbon ratios:*
**G$_{1:4}$** [K:C $\frac{1}{4}$]; **G$_{1:8}$** [K:C $\frac{1}{8}$], **G$_{1:24}$** [K:C $\frac{1}{24}$]

In an argon filled glove box (<0.1 ppm oxygen; <0.1 ppm H$_2$O), 12.00 mg (1.000 mmol) spherical graphite (SGN18) and the respective amount of potassium – **G$_{1:4}$** 9.775 mg (2.50*10$^{-1}$ mmol), **G$_{1:8}$** 4.887 mg (1:25*10$^{-1}$ mmol), **G$_{1:24}$** 1.222 mg (3.125*10$^{-2}$ mmol) – were heated under occasional stirring (spatula) at 150 °C for 8 hours. Afterwards, the salt was allowed to cool to RT and isolated as a dark blue (**G$_{1:24}$**), bronze (**G$_{1:8}$**) or beige (**G$_{1:4}$**) material, respectively.

*Synthesis of hydrogenated graphene:*

In an argon filled glove box (<0.1 ppm oxygen; <0.1 ppm H$_2$O), the graphite intercalation compound **G$_{1:4}$** (SGN18) (21.7 mg, 1 mmol carbon) was dispersed in 20 mL THF$_{(abs)}$ by the aid of a 5 min tip ultrasonication treatment (Bandelin UW 3200). Outside the glovebox under a constant Ar flow the corresponding hydrogen source (see table ST3) was added to the respective dispersion, sonicated for 3 min in a sonication bath and stirred for 12 hours. Afterwards, 5 mL water were added under a constant Ar flow and the reaction mixture was transferred into a separation funnel with 10 ml of cyclohexane. The phases were separated and the organic layer, containing the functionalized material, was purged three times with distilled water. The organic layer was filtered through a 0.2 µm reinforced cellulose membrane filter (Sartorius) and washed with water and *iso*-propanol. The covalently functionalized material was dried in vacuum.

| sample | Graphite | | | potassium | | deuteration reagent | | | |
|---|---|---|---|---|---|---|---|---|---|
| | | m (mg) | n (C) (mmol) | m (mg) | n (mmol) | | m (mg) | V (mL) | n (mmol) |
| $G_{1:4}A_H$ | SGN18 | 12 | 1 | 9.80 | 0.25 | A: $H_2O$ | 180.0 | 0.18 | 10 |
| $G_{1:4}B_H$ | SGN18 | 12 | 1 | 9.80 | 0.25 | B: MeOH | 320.4 | 0.41 | 10 |
| $G_{1:4}C_H$ | SGN18 | 12 | 1 | 9.80 | 0.25 | C: *t*-BuOH | 741.2 | 0.96 | 10 |
| $G_{1:8}A_H$ | SGN18 | 12 | 1 | 4.90 | 0.125 | A: $H_2O$ | 180.0 | 0.18 | 10 |
| $G_{1:8}B_H$ | SGN18 | 12 | 1 | 4.90 | 0.125 | B: MeOH | 320.4 | 0.41 | 10 |
| $G_{1:8}C_H$ | SGN18 | 12 | 1 | 4.90 | 0.125 | C: *t*-BuOH | 741.2 | 0.96 | 10 |
| $G_{1:24}A_H$ | SGN18 | 12 | 1 | 1.63 | 0.042 | A: $H_2O$ | 180.0 | 0.18 | 10 |
| $G_{1:24}B_H$ | SGN18 | 12 | 1 | 1.63 | 0.042 | B: MeOH | 320.4 | 0.41 | 10 |
| $G_{1:24}C_H$ | SGN18 | 12 | 1 | 1.63 | 0.042 | C: *t*-BuOH | 741.2 | 0.96 | 10 |

**Table ST1: Starting graphite intercalation compounds with varying potassium contents and different hydrogen sources used.**

*Synthesis of deuterated graphene:*

In an argon filled glove box (<0.1 ppm oxygen; <0.1 ppm $H_2O$), 1 mmol of the respective graphite intercalation compound $G_{1:n}$ (SGN18) was dispersed in 20 mL $THF_{(abs)}$ by the aid of a 5 min tip ultrasonication treatment (Bandelin UW 3200, 20 J/min, puls-rate 1s). Outside the glovebox, under a constant Ar flow, the corresponding deuterium source (see table ST1) was added to the respective dispersion, sonicated for 3 min in a sonication bath (Branson 2510, 100 W, 42kHz) and subsequently stirred for 12 hours. Afterwards, 5 mL of heavy water was added under a constant Ar flow and the reaction mixture was transferred into a separation funnel with 10 mL of cyclohexane. The phases were separated and the organic layer containing the functionalized material was purged three times with distilled water. The organic layer was filtered through a 0.2 μm reinforced cellulose membrane filter (Sartorius) and washed with water and *iso*-propanol. The covalently functionalized material was dried in vacuum.

| sample | Graphite | | | potassium | | deuteration reagent | | | |
|---|---|---|---|---|---|---|---|---|---|
| | | m (mg) | n (C) (mmol) | m (mg) | n (mmol) | | m (mg) | V (mL) | n (mmol) |
| $G_{1:4}A_D$ | SGN18 | 12 | 1 | 9.80 | 0.25 | A: $D_2O$ | 190.0 | 0.19 | 10 |
| $G_{1:4}B_D$ | SGN18 | 12 | 1 | 9.80 | 0.25 | B: MeOD | 330.4 | 0.42 | 10 |
| $G_{1:4}C_D$ | SGN18 | 12 | 1 | 9.80 | 0.25 | C: t-BuOD | 741.2 | 0.94 | 10 |
| $G_{1:8}A_D$ | SGN18 | 12 | 1 | 4.90 | 0.125 | A: $D_2O$ | 190.0 | 0.19 | 10 |
| $G_{1:8}B_D$ | SGN18 | 12 | 1 | 4.90 | 0.125 | B: MeOD | 330.4 | 0.42 | 10 |
| $G_{1:8}C_D$ | SGN18 | 12 | 1 | 4.90 | 0.125 | C: t-BuOD | 741.2 | 0.94 | 10 |
| $G_{1:24}A_D$ | SGN18 | 12 | 1 | 1.63 | 0.042 | A: $D_2O$ | 190.0 | 0.19 | 10 |
| $G_{1:24}B_D$ | SGN18 | 12 | 1 | 1.63 | 0.042 | B: MeOD | 330.4 | 0.42 | 10 |
| $G_{1:24}C_D$ | SGN18 | 12 | 1 | 1.63 | 0.042 | C: t-BuOD | 741.2 | 0.94 | 10 |

**Table ST2: Starting graphite intercalation compounds with varying potassium contents and different deuterium sources used.**

*Variation of the graphite source for the synthesis of deuterated graphene with t-BuOD as deuteration reagent:*

The synthesis was carried out as described above. The resepective materials are summarized in Table ST2.

| sample | Graphite | | | potassium | | deuteration reagent | | | |
|---|---|---|---|---|---|---|---|---|---|
| | | m (mg) | n (C) (mmol) | m (mg) | n (mmol) | | m (mg) | V (mL) | n (mmol) |
| $G_{1:4}C_D$ | SGN18 | 12 | 1 | 9.80 | 0.25 | C: t-BuOD | 741.2 | 0.94 | 10 |
| $G_{1:4}C_D$ | PEX10 | 12 | 1 | 9.80 | 0.25 | C: t-BuOD | 741.2 | 0.94 | 10 |
| $G_{1:4}C_D$ | NG | 12 | 1 | 9.80 | 0.25 | C: t-BuOD | 741.2 | 0.94 | 10 |

**Table ST3: Variation of the graphite starting material: SGN18, PEX10, natural graphite (NG).**

*Synthesis of **G$_{1:4}$A$_D$(NH$_3$)**:*

In a 100 mL round bottom flask, 30 mL of liquid ammonia were condensed at -70 °C and the graphite intercalation compound **G$_{1:4}$** (SGN18) (21.7 mg, 1 mmol carbon) was added under a constant Ar flow. After a blue coloring of the solution 0.19 mL of D$_2$O (10 eq.) was added to the respective dispersion. The system was stirred for 2 hours and subsequently the ammonia was evaporated by warming the reaction mixture to rt. Afterwards, 5 mL of D$_2$O were added under a constant Ar flow and the reaction mixture was transferred into a separation funnel with 10 mL of cyclohexane. The phases were separated and the organic layer, containing the functionalized material, was purged three times with distilled water. The organic layer was filtered through a 0.2 µm reinforced cellulose membrane filter (Sartorius) and washed with water and *iso*-propanol. The covalently functionalized material was dried in vacuum.

*Synthesis of **G$_{1:8}$A$_D$(solvent free)**:*

Under a constant Ar flow 3 mL D$_2$O were added drop-wise to the potassium intercalation compound **G$_{1:8}$** (SGN18) (16.8 mg, 1 mmol carbon) at 0 °C and stirred for 12 hours. Afterwards, 5 mL of D$_2$O were added under a constant Ar flow and the reaction mixture was transferred into a separation funnel with 10 mL of cyclohexane. The phases were separated and the organic layer, containing the functionalized material, was purged three times with distilled water. The organic layer was filtered through a 0.2 µm reinforced cellulose membrane filter (Sartorius) and washed with water and *iso*-propanol. The covalently functionalized material was dried in vacuum.

## 4. Additional Analytical Data

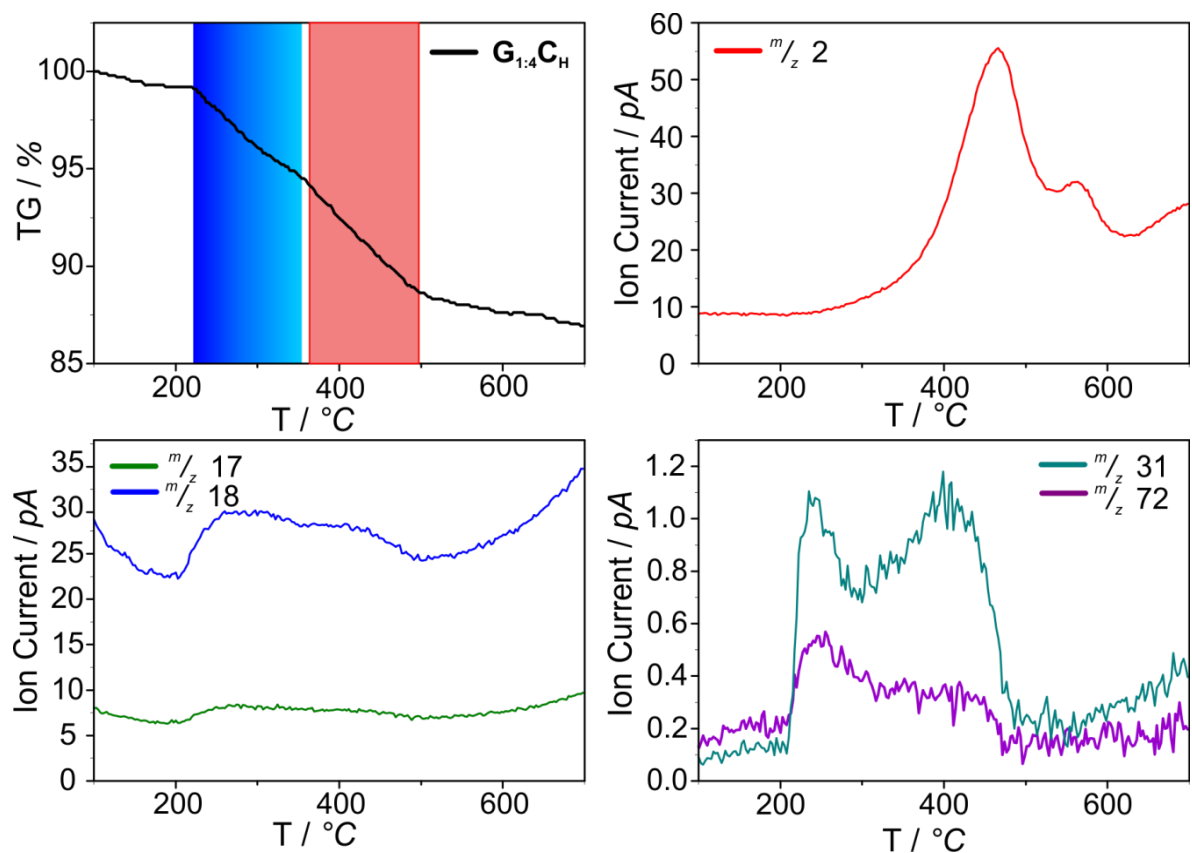

**Figure S1:** Top left: TG-profile of hydrogenated graphene **G$_{1:4}$C$_H$**. Top right: MS traces for $^m/_z$ 2 (H$_2$). Bottom line: MS profiles for $^m/_z$ 17, 18 (H$_2$O, OH), and 31, 72 (THF).

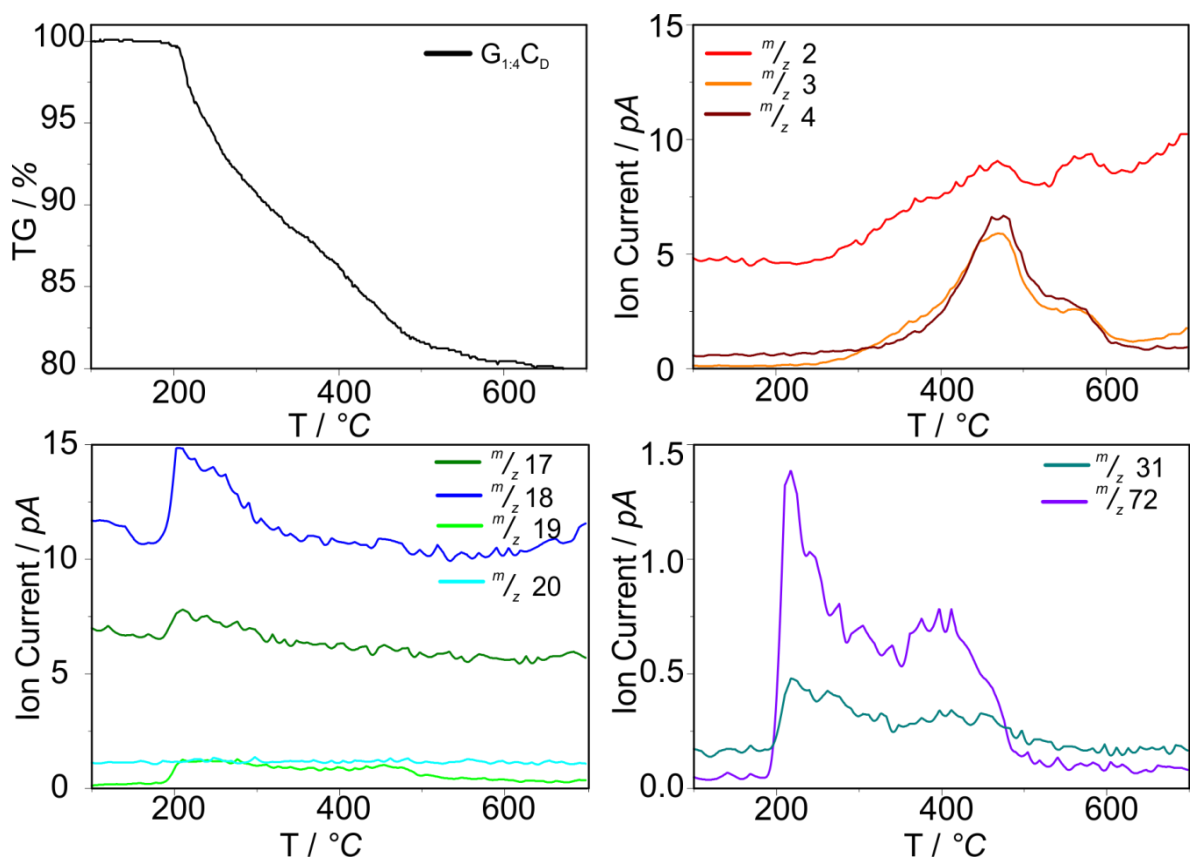

**Figure S2:** Top left: TG-profile of deuterated graphene **G$_{1:4}$C$_D$**. Top right: MS traces for $^m/_z$ 2, 3, 4 (H$_2$, HD, D$_2$). Bottom line: MS profiles for $^m/_z$ 17, 18, 19, 20 (H$_2$O, OH, OD, D$_2$O), and 31, 72 (THF).

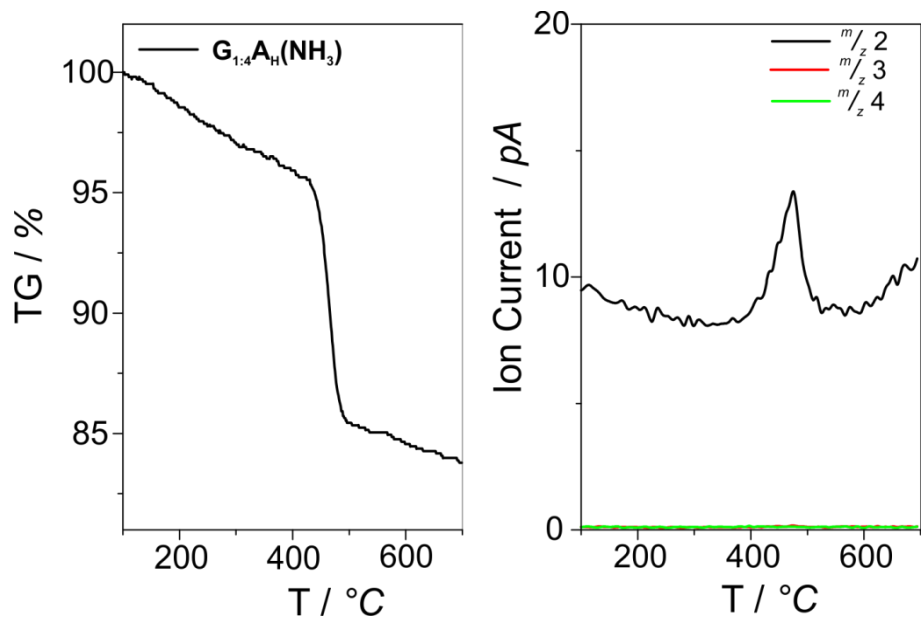

**Figure S3:** Left: TG profile of deuterated graphene **G$_{1:4}$A$_H$(NH$_3$)**. Right: MS traces for $m/z$ 2, 3, 4 (H$_2$, HD, D$_2$).

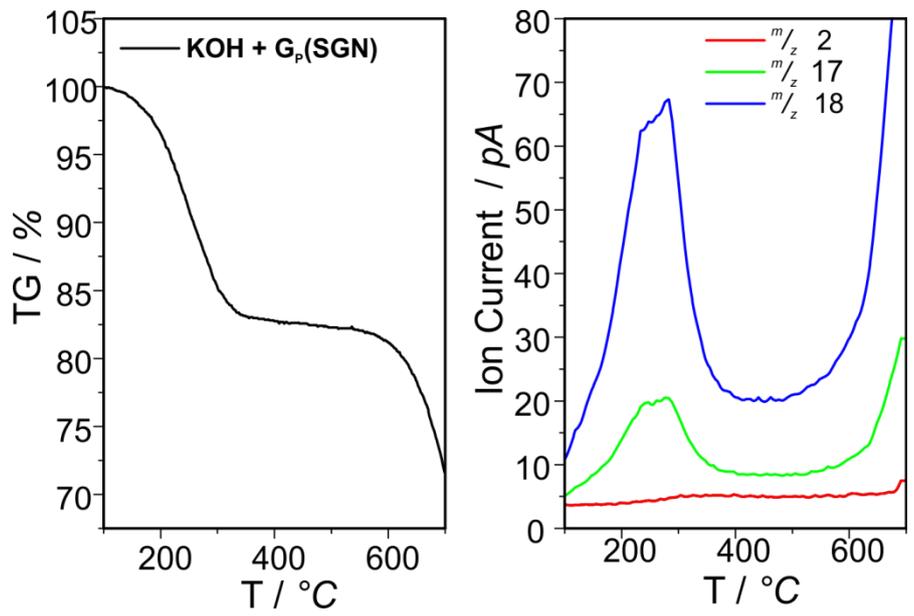

**Figure S4:** Left: TG profile of KOH+ **G$_P$(SGN)**. Right: MS traces for $m/z$ 2 (H), and 17, 18 (H$_2$O, OH).

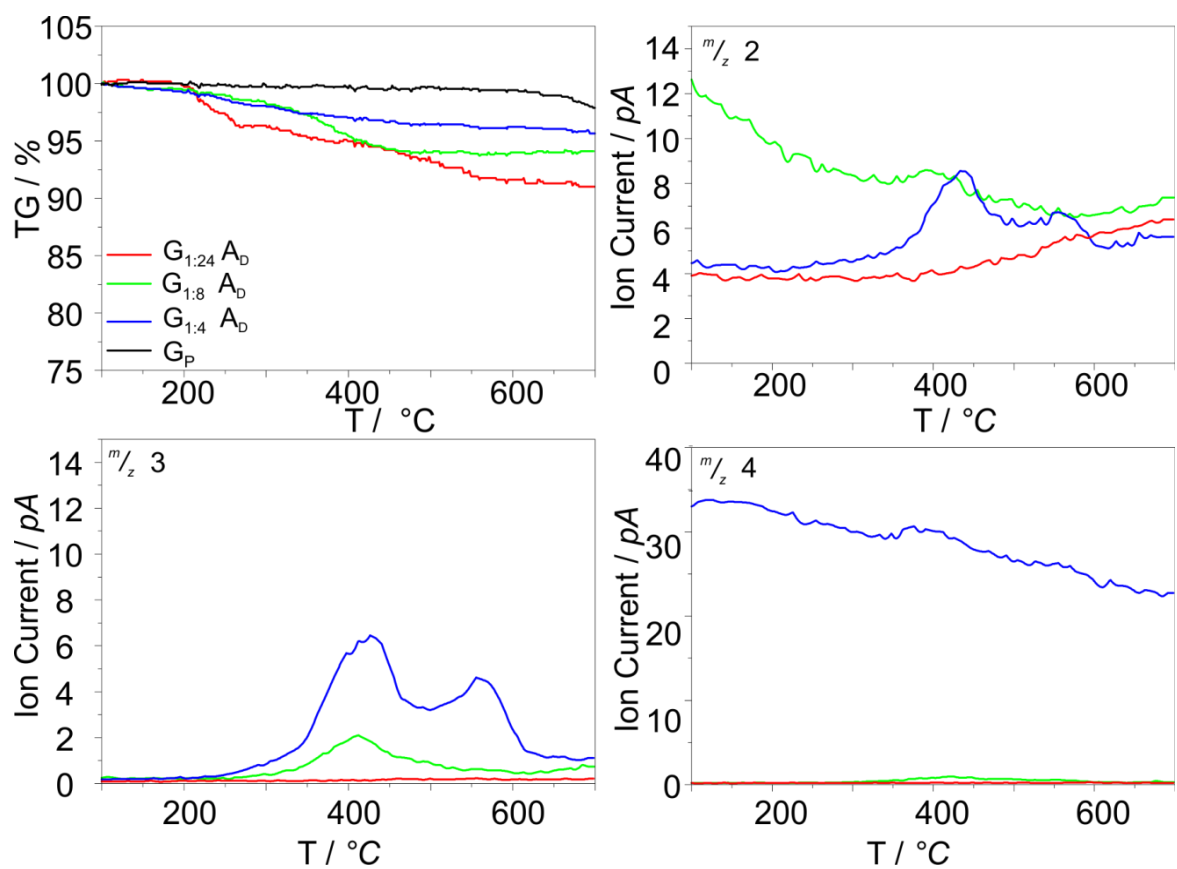

**Figure S5:** Top left: TG-profiles of deuterated graphenes **G$_{1:4}$A$_D$**, **G$_{1:8}$A$_D$**, and **G$_{1:24}$A$_D$**, variation of the potassium content, D$_2$O as deuterium source. Top right and bottom line: Corresponding MS traces for *m/z* 2, 3, 4 (H$_2$, HD, D$_2$).

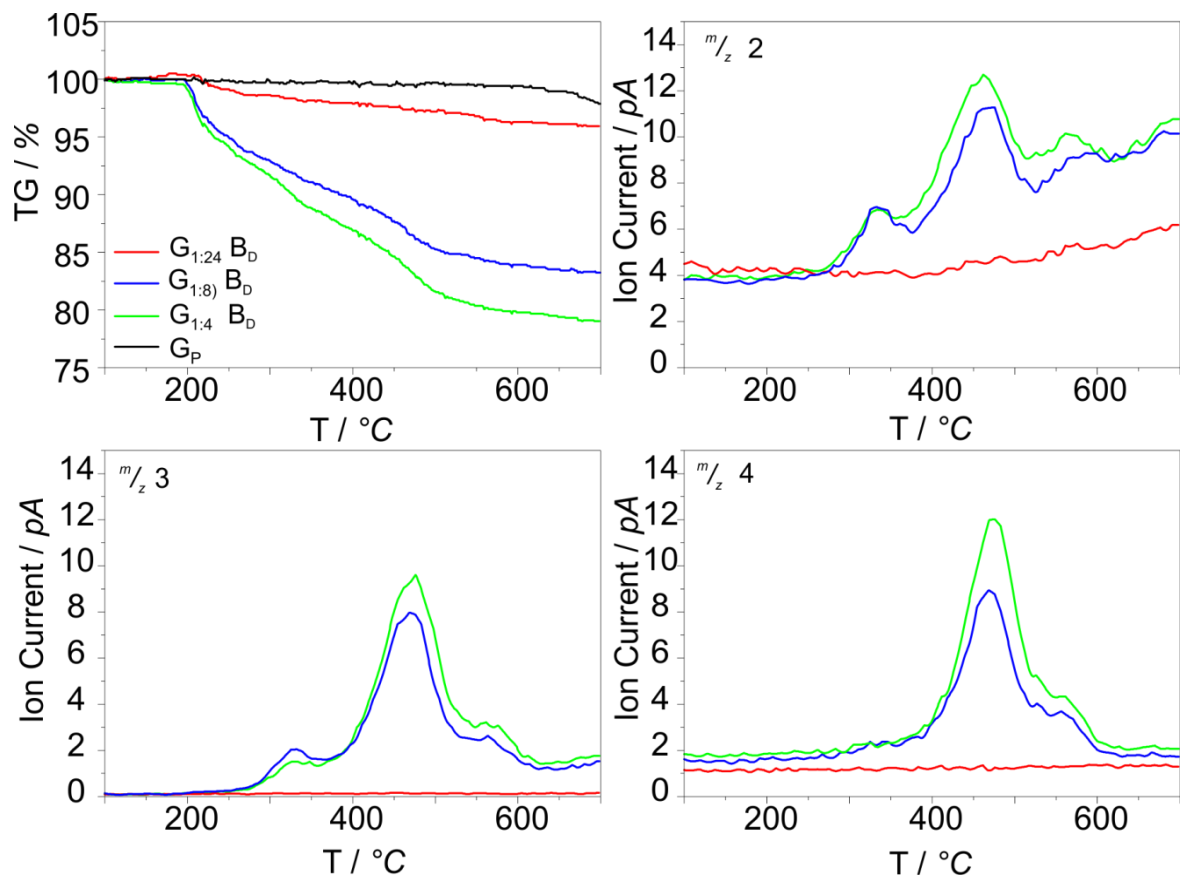

**Figure S6:** Top left: TG profiles of deuterated graphenes **G$_{1:4}$B$_D$**, **G$_{1:8}$B$_D$**, and **G$_{1:24}$B$_D$**, variation of the potassium content, MeOD as deuterium source. Top right and bottom line: Corresponding MS traces for $^m/_z$ 2, 3, 4 (H$_2$, HD, D$_2$).

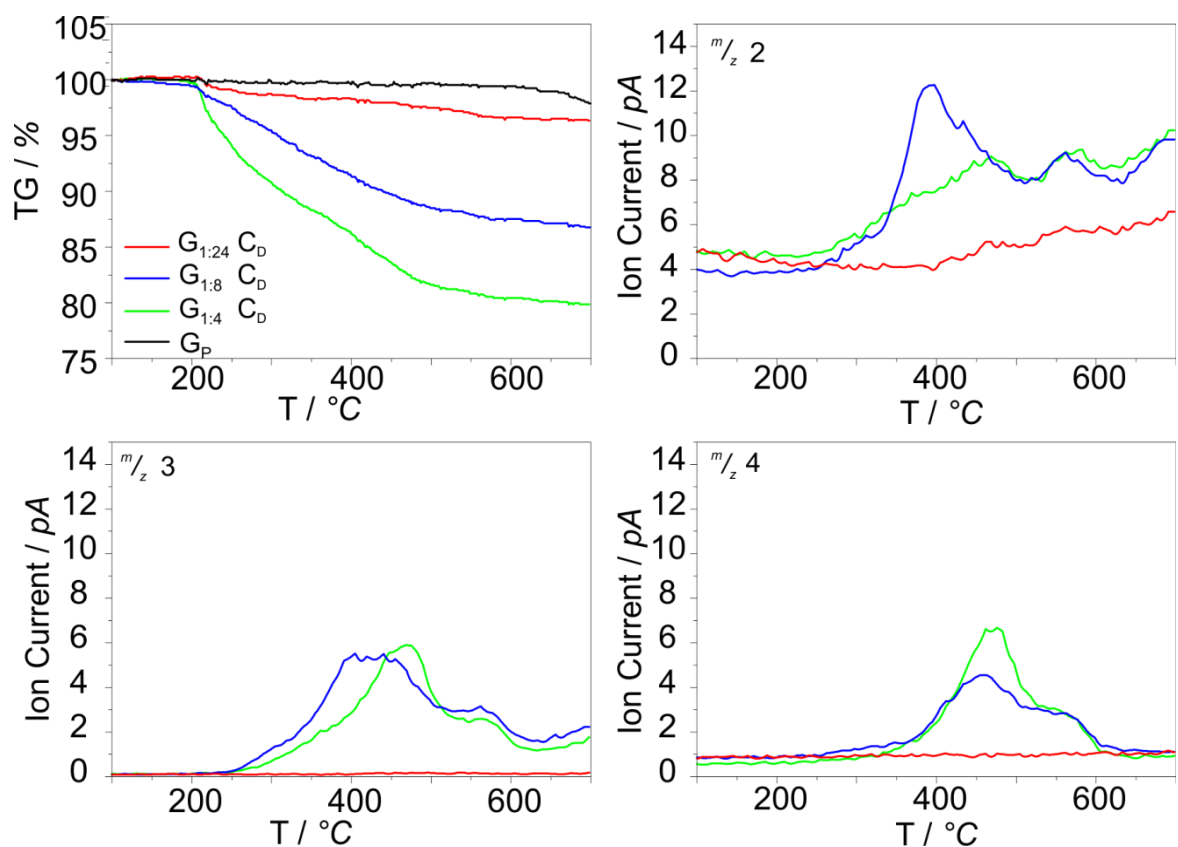

**Figure S7:** Top left: TG profiles of deuterated graphenes **G$_{1:4}$C$_D$**, **G$_{1:8}$C$_D$**, and **G$_{1:24}$C$_D$**, variation of the potassium content, *t*-BuOD as deuterium source. Top right and bottom line: Corresponding MS traces for $^m/_z$ 2, 3, 4 (H$_2$, HD, D$_2$).

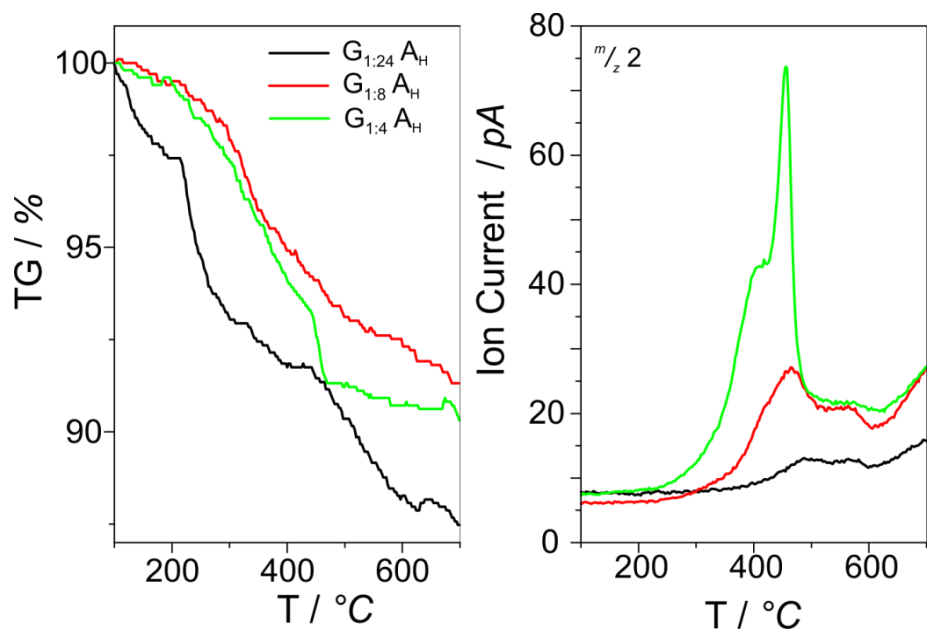

**Figure S8:** Left: TG profiles of hydrogenated graphenes **G$_{(1:4)}$A$_H$, G$_{(1:8)}$ A$_H$**, and **G$_{(1:24)}$ A$_H$**, variation of the potassium content, H$_2$O as proton source. Right: Corresponding MS traces for $^m/_z$ 2 for H$_2$.

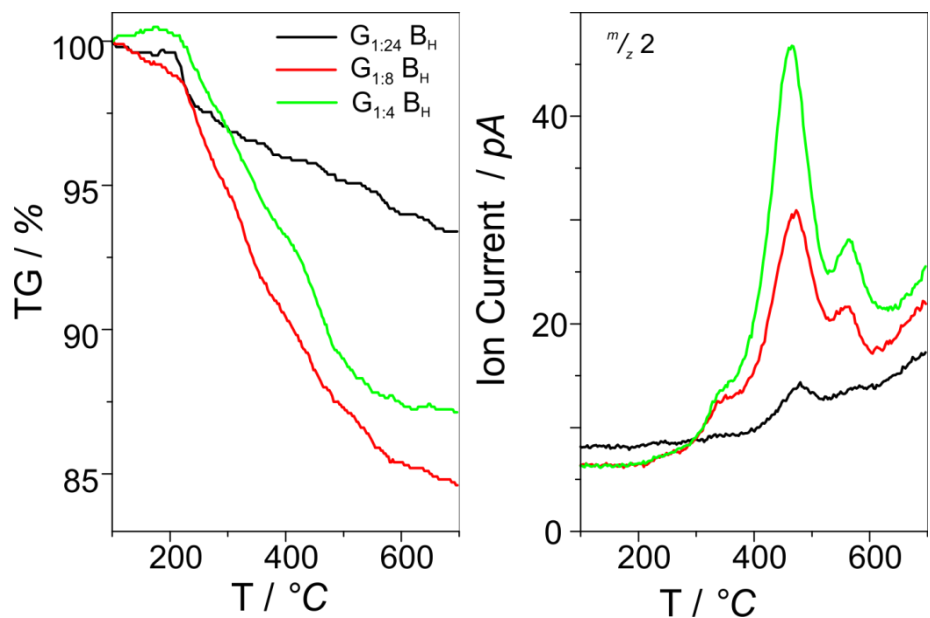

**Figure S9:** Left: TG profiles of hydrogenated graphenes **G$_{1:4}$B$_H$, G$_{1:8}$B$_H$**, and **G$_{1:24}$B$_H$**, variation of the potassium content, MeOH as proton source. Right: Corresponding MS traces for $^m/_z$ 2 for H$_2$.

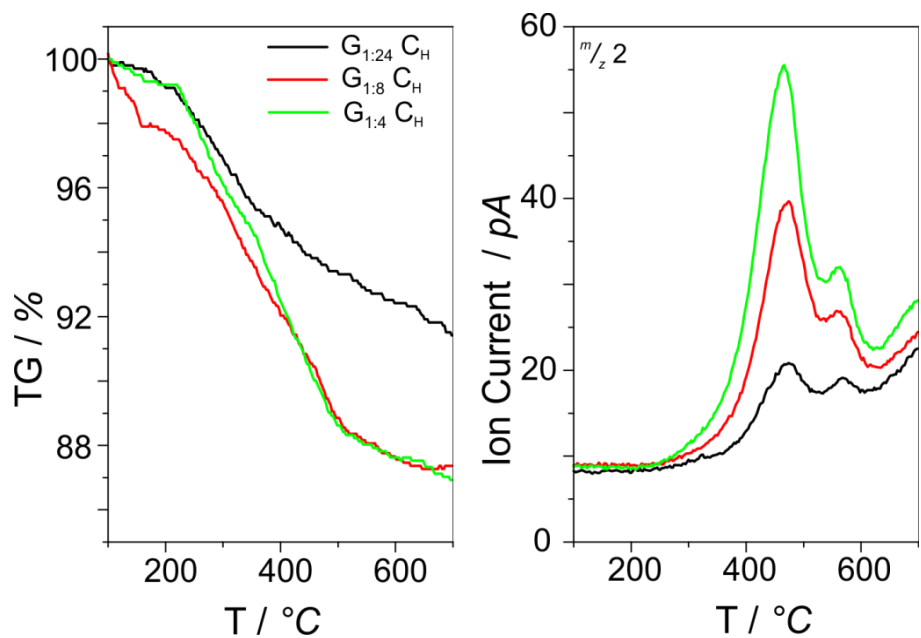

**Figure S10:** Left: TG profiles of hydrogenated graphenes **G$_{1:4}$C$_H$, G$_{1:8}$C$_H$**, and **G$_{1:24}$C$_H$**, variation of the potassium content, *t*-BuOH as proton source. Right: Corresponding MS traces for $^m/_z$ 2 for H$_2$.

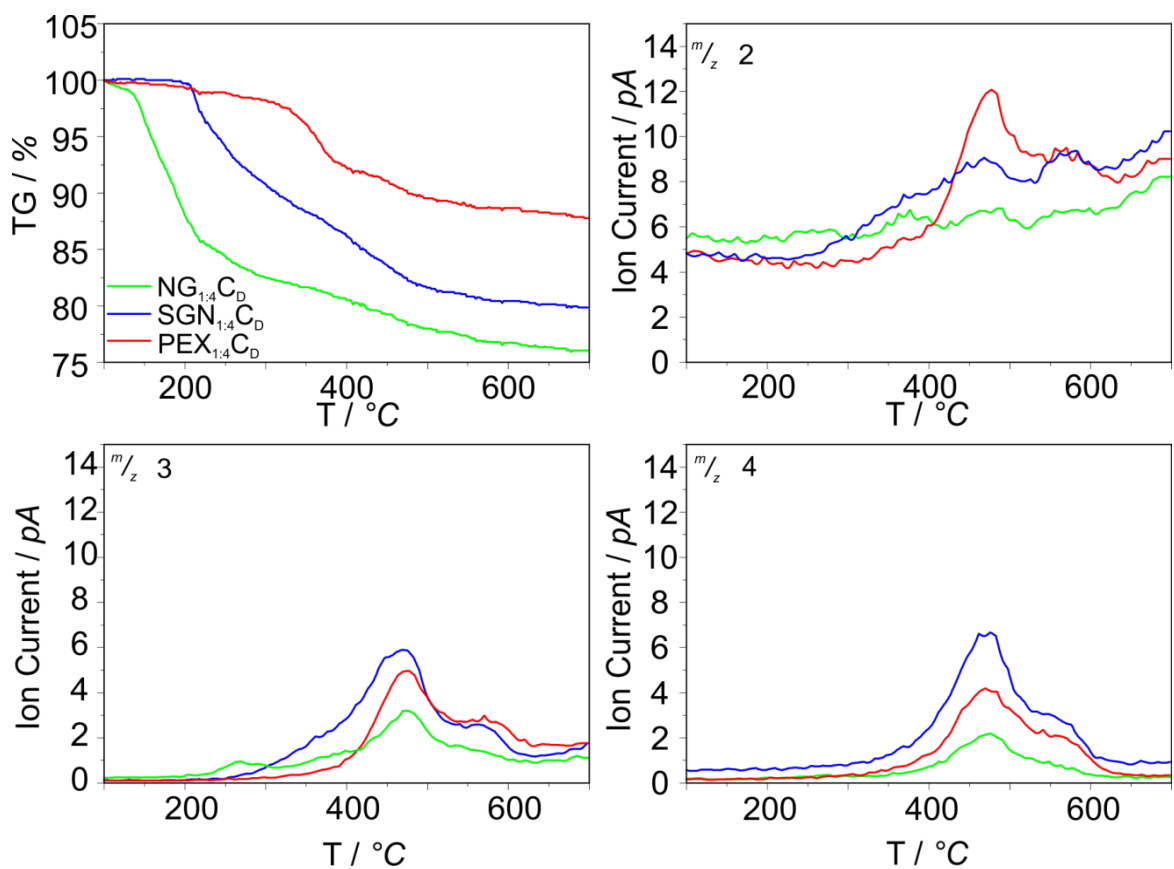

**Figure S11:** Top left: TG profiles of deuterated graphenes **NG$_{1:4}$C$_D$**, **SGN$_{1:4}$C$_D$**, and **PEX$_{1:4}$C$_D$**, *t*-BuOD as deuterium source. Top right and bottom line: Corresponding MS traces for $^m/_z$ 2, 3, 4 (H$_2$, HD, D$_2$).

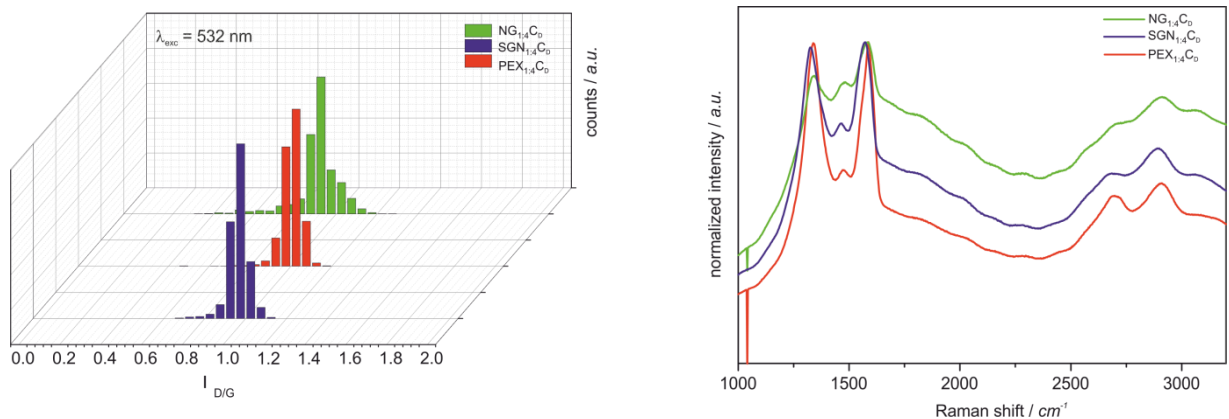

**Figure S12**: Left: Statistical Raman $I_{D/G}$ intensity distribution of **NG$_{1:4}$C$_D$**, **SGN$_{1:4}$C$_D$**, and **PEX$_{(1:4)}$C$_D$**, variation of the starting graphite, *t*-BuOD as deuterium source. Right: Corresponding Raman mean spectra - $\lambda_{exc.}$ = 532 nm.

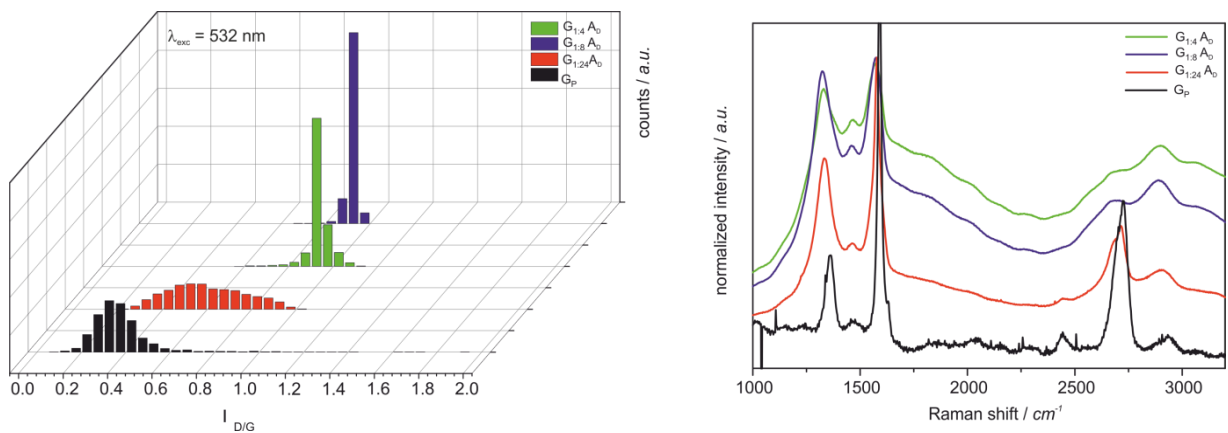

**Figure S13**: Left: Statistical Raman $I_{D/G}$ intensity distribution of $G_P$, $G_{1:4}A_D$, $G_{1:8}A_D$, and $G_{1:24}A_D$, variation of the potassium content, $D_2O$ as deuterium source. Right: Corresponding Raman mean spectra - $\lambda_{exc.}$ = 532 nm.

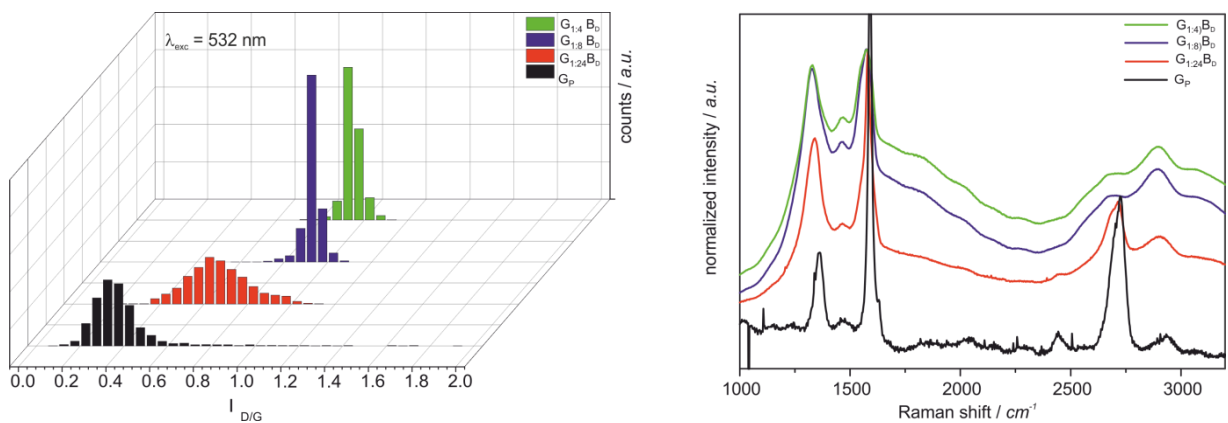

**Figure S14**: Left: Statistical Raman $I_{D/G}$ intensity distribution of $G_P$, $G_{1:4}B_D$, $G_{1:8}B_D$, and $G_{1:24}B_D$, variation of the potassium content, MeOD as deuterium source. Right: Corresponding Raman mean spectra - $\lambda_{exc.}$ = 532 nm.

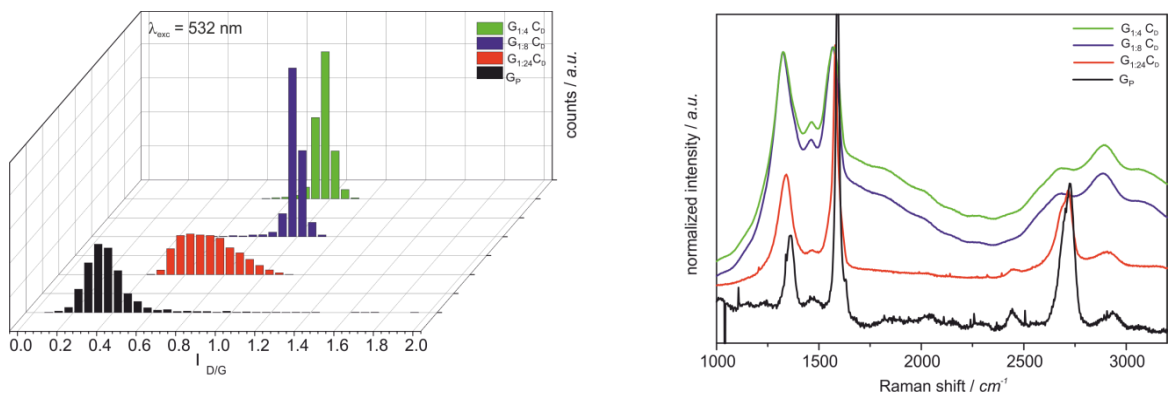

**Figure S15**: Left: Statistical Raman $I_{D/G}$ intensity distribution of $G_P$, $G_{1:4}C_D$, $G_{1:8}C_D$, and $G_{1:24}C_D$, variation of the potassium content, $t$-BuOD as deuterium source. Right: Corresponding Raman mean spectra - $\lambda_{exc.}$ = 532 nm.

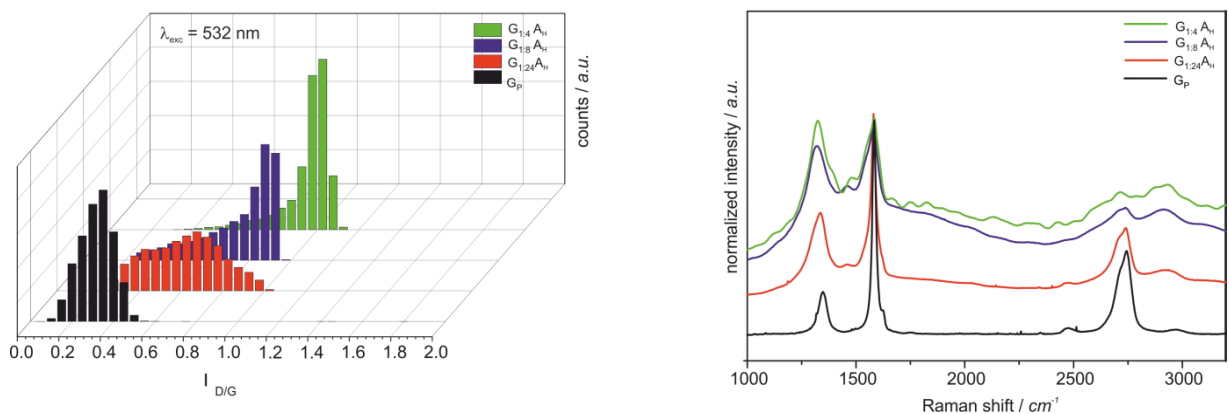

**Figure S16**: Left: Statistical Raman $I_{D/G}$ intensity distribution of $G_P$, $G_{1:4}A_H$, $G_{1:8}A_H$, and $G_{1:24}A_H$, variation of the potassium content, $H_2O$ as proton source. Right: Corresponding Raman mean spectra - $\lambda_{exc.}$ = 532 nm.

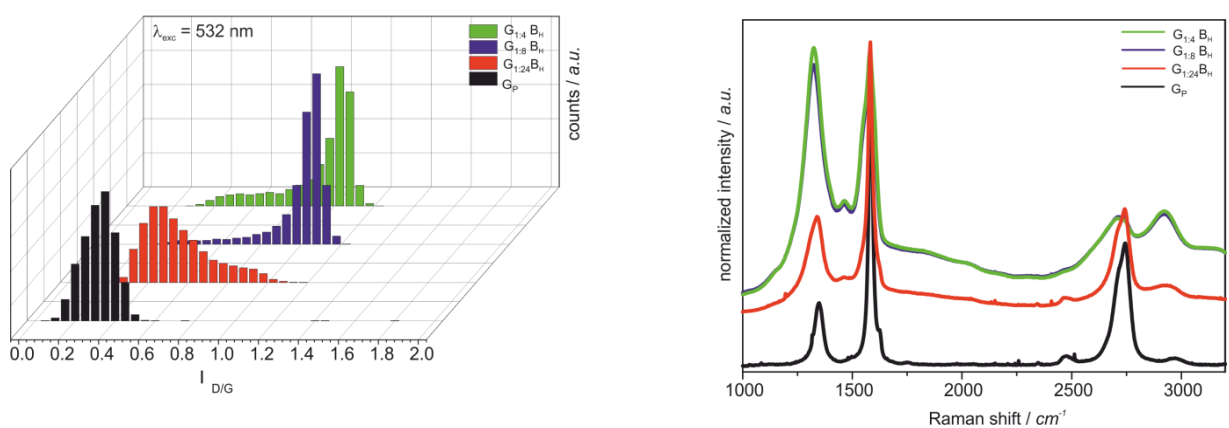

**Figure S17**: Left: Statistical Raman $I_{D/G}$ intensity distribution of $G_P$, $G_{1:4}B_H$, $G_{1:8}B_H$, and $G_{1:24}B_H$, variation of the potassium content, MeOH as proton source. Right: Corresponding Raman mean spectra - $\lambda_{exc.}$ = 532 nm.

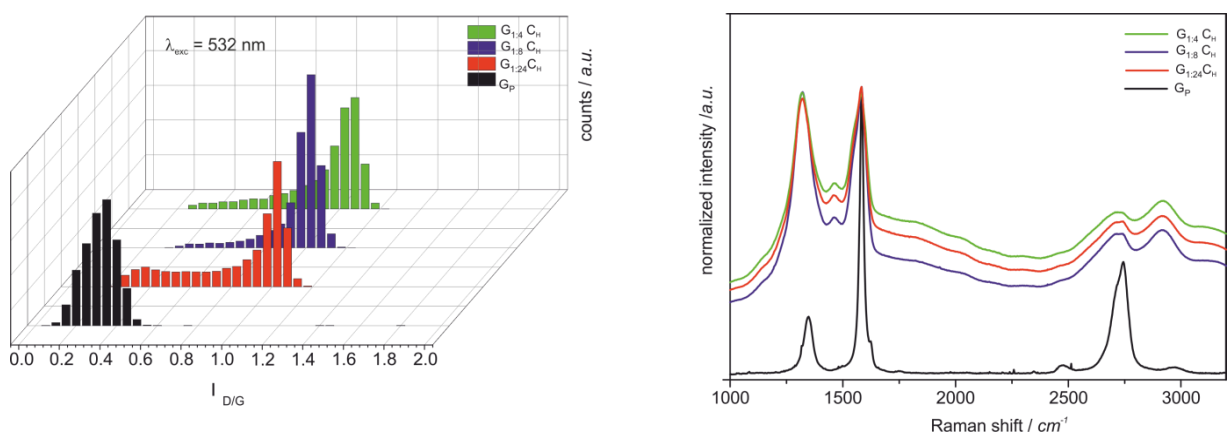

**Figure S18**: Left: Statistical Raman $I_{D/G}$ intensity distribution of $G_P$, $G_{1:4}C_H$, $G_{1:8}C_H$ and $G_{1:24}C_H$, variation of the potassium content, *t*-BuOH as proton source. Right: Corresponding Raman mean spectra - $\lambda_{exc.}$ = 532 nm. .